# Key components for nano-assembled plasmon-excited single molecule non-linear devices

**Günter Kewes[1,*], Max Schoengen[2], Giacomo Mazzamuto[3], Oliver Neitzke[1], Rolf-Simon Schönfeld[1], Andreas W. Schell[1], Jürgen Probst[2], Janik Wolters[1], Bernd Löchel[2], Costanza Toninelli[3,4], and Oliver Benson[1]**

[1]*Institut für Physik, AG Nano-Optik, Humboldt-Universität zu Berlin, Newtonstraße 15, Germany*
[2]*Helmholtz-Zentrum Berlin für Materialien und Energie, Albert-Einstein-Straße 15, 12489 Berlin, Germany*
[3]*European Laboratory for Non-Linear Spectroscopy (LENS), University of Florence, Via Nello Carrara 1, 50019 Sesto Fiorentino Florence, Italy*
[4]*INO, Istituto Nazionale di Ottica, Largo Fermi 6, 50125 Firenze, Italy*
[*]*gkewes@physik.hu-berlin.de*

**Abstract:** Tremendous enhancement of light-matter interaction in plasmon-excited molecular hybrid devices allows for non-linearities on the level of single emitters and few photons. This promises a plethora of novel applications like single photon transistors. Nevertheless, building the components of such devices is technologically extremely challenging. We tackle this task by lithographically fabricating on-chip plasmonic waveguides, efficiently connected to far-field in- and out-coupling ports via low-loss dielectric waveguides. Furthermore, a nano-assembling technology is developed, enabling the controlled coupling of single organic emitters to the plasmonic waveguides. Dibenzoterrylene fluorescent molecules hosted in anthracene crystals are investigated for this purpose. Here we present all key-components and technologies for a plasmon-excited single molecule non-linear device.

Transistors are based on non-linear responses that enable signal control by means of a weak gating signal. While in electronics strong electron-electron interaction allowed continuous improvement leading to extremely efficient transistors working with low switching currents, non-linear optical devices suffer from very low non-linear coefficients and thus require high-intensity laser beams. Nevertheless, in recent years promising concepts have been proposed and realized [1–5] that exploit individual quantum systems as non-linear elements operating even on the few to single photon level. In order to achieve the desired non-linear response – ideally in an integrated device – a sufficiently strong coupling between single emitters and confined light is mandatory.

The interaction strength can be enhanced by using emitters with a large transition dipole moment, or by resonantly enhancing and tightly confining the electromagnetic field. Dielectric micro-resonators [6–9] or localized plasmonic modes [10,11] have been used for this purpose.

Complementary to all-integrated approaches [12], assembly of hybrid devices allow for an independent optimization of the quantum emitters and the light guiding structures [13]. Within these architectures, plasmonic waveguides have the advantage that the energy density in the surface plasmon polariton is more strongly increased compared to purely photonic modes [14]. In addition, plasmonic enhancement does not rely on narrow resonances and thus allows a larger bandwidth of the non-linear devices [3]. Theoretical proposals predict switches working at the single plasmon level [3] and even strong coupling of emitters to plasmonic modes [10].

For an experimental realization of a single photon non-linear device such as the single photon transistor proposed by Chang et al. [3], there are three key ingredients:

i) an integrated dielectric-plasmonic coupler-structure with high coupling efficiency,
ii) a stable single quantum emitter with predictable properties and large optical dipole moment,
iii) an assembling approach to bring the first two components together.

In this paper, we demonstrate all these three challenging requirements.

Figure 1 illustrates schematically our device architecture. A Bragg-grating coupler (not shown) couples far-field photons into a dielectric $Si_3N_4$ waveguide. Then, an efficient V-type photon-to-plasmon transducer (details of its design and numerical simulation in Ref. [15]) transfers excitation into a short plasmon-waveguide with tight confinement. There, a single organic dibenzoterrylene (DBT) molecule precisely positioned next to the plasmonic waveguide will scatter or transmit the incoming plasmons depending on its internal states, yielding a non-linear behavior. Finally, plasmons are converted into guided photons and eventually coupled into free-space again.

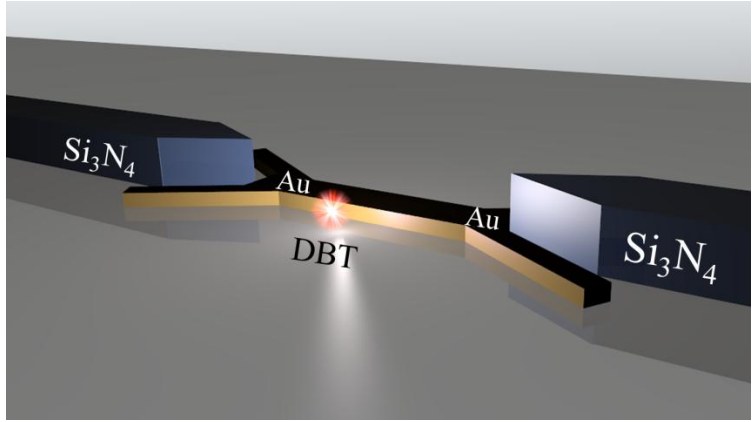

Fig. 1. Scheme of a single photon non-linear device: A dielectric waveguide made from $Si_3N_4$ (blueish color) guide photons to a V-shaped photon-to-plasmon transducer, where plasmons in a short plasmonic gold stripe waveguide (golden color) are excited. A precisely aligned emitter (DBT molecule) close to the plasmonic waveguide intensely interacts with the plasmons. After the interaction plasmons are transduced back to photons and guided away by a second dielectric waveguide.

In the following, we introduce our approaches and describe in detail their experimental realization.

*i) An integrated dielectric-plasmonic coupler*

The first requirement is to route photons with high efficiency to the active zone of the non-linear device. This zone consists of a single DBT molecule interacting with guided plasmons. The strong field enhancement guarantees that the emitter's decay rate $\Gamma_{tot}$ is dominated by the emission into a single guided plasmon mode with rate $\Gamma_{pl}$ [10]. In a realistic scenario the coupling to weakly guided or radiating modes [16] and the presence of more than one bound mode in planar waveguides [17] has to be taken into account. At the same time, the coupling strength to the plasmonic mode has to be balanced with the unavoidable propagation losses. Therefore, the plasmonic waveguide has to be as short as possible. This can be obtained by guiding excitations mainly through dielectric waveguides, since they exhibit much lower or even negligible losses. Unavoidable loss is caused by the dielectric-to-plasmon transducer.

However, as we will show, an optimized transducer geometry (see Ref. [15]) allows for very high single mode conversion. Finally, coupling into (and out of) the integrated dielectric waveguides can be obtained by Bragg-grating far-field couplers.

We fabricated an elaborate dielectric-plasmonic chip. The photonic part is built from a 300 nm layer of low pressure chemical vapor deposition (LPCVD) $Si_3N_4$, grown on a thermally oxidized silicon substrate (Silicon Valley Microelectronics, Inc.). The $Si_3N_4$ is almost absorption free in the desired wavelength range (785 nm). For the plasmonic part of the chip we use a bi-layer consisting of 5 nm Cu and 60 nm Au, evaporated on the chip. This material system represents a compromise between fabrication constraints, stability under ambient conditions and absorption losses by Ohmic resistance in the metallic structure.

To produce the combined photonic-plasmonic structures two complementary lithographic techniques are used. The $Si_3N_4$ can only be structured by a subtractive method [18], i.e. a positive etch-mask is fabricated and the surrounding nitride is removed via etching. In contrast, the plasmonic gold parts have to be built up via additive nanostructuring. Both methods need etch masks in opposite tones. In addition the additive structuring needs a sacrificial layer for a lift-off process. These circumstances and the gap between the dielectric and plasmonic part in the nanometer range impede a separation of the structuring steps.

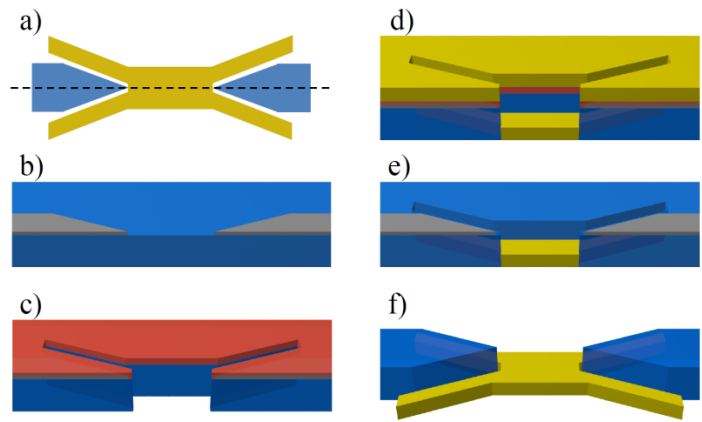

Fig. 2. Simplified overview of the different steps during processing of the photonic-plasmonic on-chip structures. b)-e) show a cut along the dashed line in a).
b) Ni mask (grey) for the dielectric waveguides in $Si_3N_4$ (blue),
c) ZEP resist (red) / $Si_3N_4$ mask for the plasmonic components after the second etching step,
d) gold evaporation (yellow layer comprises Cu, Au and Ni layers), e) lift-off of gold,
f) final etching step (and removal of remaining Ni on gold structures).

Figure 2 gives an overview of the fabrication process. For the electron beam lithography (EBL) steps we use a Vistec 5000plus system with 100 kV acceleration voltage, an Oxford plasmalab 80 plus for reactive ion etching (RIE) and a Leybold thermal evaporation system.

In the first EBL step, square markers with an edge length of 20 µm are fabricated using a Ti-Au-Ni-layer system (not shown in figure 2). Later on, these markers allow alignment of the three subsequent EBL steps with an accuracy of at least 5 nm.

In the second EBL step, the coupler gratings of the dielectric grating couplers (consisting of coupler grating and reflection grating) are fabricated (not shown in figure 2). Therefore a PMMA (2.2 M, Microchemicals) resist layer is used as an etch mask and the structure is transferred to the substrate via a highly anisotropic $CHF_3$-RIE-process (etch depth 60 nm).

In a third EBL step a nickel etch mask for the dielectric waveguides and the reflecting gratings (Bragg mirrors) of the grating couplers are prepared. Therefore the recoated substrate (PMMA 2.2 M) is exposed by EBL and the resulting positive mask is evaporated with 10 nm Ni. The subsequent lift-off step is done using dimethylformamide (DMF) (figure 2b)).

To achieve the required small gaps between plasmonic and dielectric waveguide, the $Si_3N_4$-layer is also used as a mask for the plasmonic parts in the fourth lithography step. Therefore the substrate is coated with a 150 nm thick ZEP 7000 (Zeon Corporation, Japan) resist layer (burying the thin Ni etch mask underneath it), which is structured with the positive tone of the plasmonic coupler. Compared to standard PMMA, the ZEP resist has a higher etch resistance, thus the resist remaining after etching is still thick enough to be used as sacrificial layer in the following lift-off process. A thicker PMMA layer cannot be used, because the required structure sizes could then not be achieved due to the limited aspect ratio. The $Si_3N_4$ layer is partially removed in a highly anisotropic $CHF_3$-RIE process at the designated positions of the plasmonic parts (figure 2c)). At this step the control of the etch depth allows adjusting of the vertical placement of the plasmonic coupler, which might be used as an additional control parameter to achieve coupling efficiencies even higher than predicted in Ref. [15]. Actually, for the structures characterized in this study, the plasmonic parts are fabricated on top of a 115 nm $Si_3N_4$ plateau of the same cross section, elevating the plasmonic waveguide to the center of the dielectric waveguide (not shown in figure 1 and 2 for simplicity).

The plasmonic Cu-Au layer system is evaporated with a high working distance of 45 cm which results in only thin walls at the edges of the plasmonic coupler (figure 2d)). The particular choice of Cu guarantees a smooth gold surface, Cu is not affected by the last wet etch process and it has relatively small Ohmic losses. Additionally, the gold is covered with a 10 nm Nickel layer (not shown in figure 2 for simplicity) for high resistance to the final $CHF_3$-RIE process.

Removal of the sacrificial ZEP layer is again done with a DMF lift-off (figure 2e)). The final process steps are the etching of the dielectric waveguides in a highly anisotropic $CHF_3$-RIE process and the removal of the Ni etch masks by wet etching with hydrochloric acid (figure 2f)). Figure 3 shows electron micrographs of the fabricated structures before Ni removal.

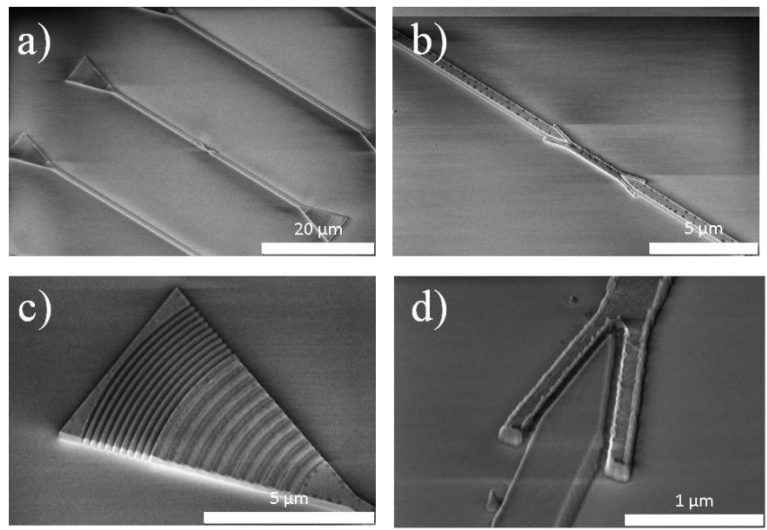

Fig. 3. Scanning electron micrographs of the fabricated structures ( a) - c) before Ni removal).
a) Overview of whole structure showing the dielectric grating couplers and the dielectric waveguides with a short plasmonic waveguide in its middle. The pure dielectric waveguides above and below are used as references.
b) Zoom-in showing the plasmonic waveguide and the dielectric-plasmonic transducers.
c) The photonic grating-coupler, consisting of a partially etched grating and a Bragg-reflector.

d) Zoom-in to the photon-to-plasmon transducer with a 40 nm gap between metallic and dielectric components.

To characterize the fabricated structures, we concentrate on the photon-to-plasmon transducer, as the used Bragg gratings are well established [19,20]. In particular the coupling efficiency η, given by the ratio of the excited plasmons to the number of photons in the waveguide, is determined. To this end, we compare the transmission of the fabricated photonic-plasmonic structures to the transmission of purely dielectric reference structures fabricated on the same sample. Note that variations in transmission through the reference structures are below 1 % indicating a high quality of the chip fabrication. In our measurements, we use a homebuilt inverted microscope with a charge-coupled device camera (CCD) for detection. One of the Bragg-gratings sitting on the end of the dielectric waveguide is fully illuminated by a weakly focused laser ($\lambda = 785$ nm). The sample is aligned to optimize the transmission to the Bragg-grating sitting at the opposite end. Images are taken by the CCD camera and analyzed (cf. Fig. 4a)). By normalizing the intensity at the output grating with respect to the flux emerging from the reference structures without plasmonic components, the overall efficiency of the plasmonic components is determined. To eliminate contributions from propagation losses in the plasmonic waveguides and to extract the efficiency η of the photon-to-plasmon transducers, the measurement is repeated for a series of waveguide structures with different plasmon-waveguide length L (Fig. 4b)): the extrapolation of the slowly decaying component of a double exponential fit ($f(z) = Ae^{-\alpha z} + Be^{-\beta z}$) to L = 0 reveals a conversion efficiency $\eta^2 = (32 \pm 12)$ % for two couplers and hence $\eta = (57 \pm 21)$ % can be deduced for a single photon-to-plasmon transducer. This value is consistent to numerical simulations (JCMwave, see Ref. [15]) of the fabricated structures predicting η = 40 %. The exponential fit in Figure 4b) matches the data points for large lengths L as it is expected for a plasmon waveguide with propagation losses leading to an exponential damping. For very short waveguide lengths the apparent coupling is larger than the estimated photon-to-plasmon conversion efficiency, since photons directly couple into the second dielectric waveguide by scattering. This is the reason for using the double-exponential fit. The value of the fitted curve's slow-decay component at zero propagation length yields the coupling efficiency. This lower decay-constant corresponds to an attenuation constant $\alpha = (0.54 \pm 0.14)$ µm$^{-1}$ or a plasmon propagation length, i.e. the distance after which the plasmon intensity decays to 1/e of the initial intensity, of $\lambda_{pl} = (1.84 \pm 0.47)$ µm. This is in agreement with simulations where $\alpha = 0.49$ µm$^{-1}$ and $\lambda_{pl} = 2.05$ µm were found, respectively.

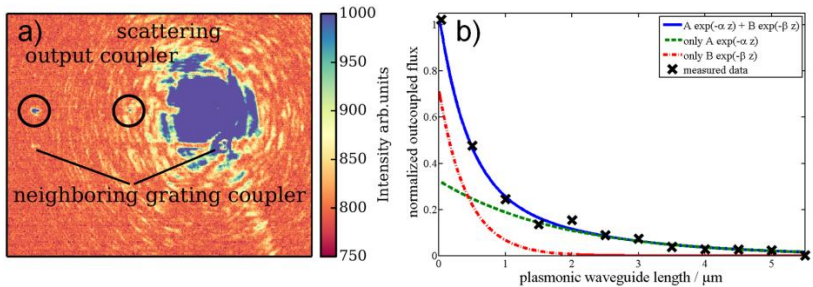

Fig. 4. a) CCD camera image corresponding to a measurement for estimating the coupling efficiency η of the photon-to-plasmon transducer. The right blue spot is a reflection from a laser scattered at the first Bragg-grating, small blue spots (both indicated by black circles) correspond to scattered photons from the photon-to-plasmon transducer and from the second Bragg-grating, respectively.
b) Measured relative out-coupled flux (crosses) as a function of plasmonic waveguide length L. The solid curve is a double exponential fit to the signal, showing that typical plasmon damping is present. Dashed lines show the contributions of the fast and slowly decaying components, respectively. Fitting the slow decay at zero waveguide length yields a coupling efficiency of $\eta = (57 \pm 21)$ %. The coupling efficiency seems to be high for short length L only due to direct coupling of the photonic waveguides by scattered photons (fitted by the fast exponential decay).

*ii) A stable single quantum emitter with predictable properties*

The second requirement concerns the single quantum emitters employed: they should have a large optical dipole moment, a high fluorescence quantum yield, lifetime limited linewidth of their optical transition, and a predictable emission wavelength to allow for matching to the fabricated dielectric- and photonic structures. For practical reasons, they should also be sufficiently stable even under ambient conditions.

Single dibenzoterrylene (DBT) fluorescent molecules hosted in anthracene (AC) crystals are almost ideal [21]: They have outstanding photostability, strong dipole transitions in the near-infrared (785 nm), and lifetime-limited linewidth at cryogenic temperatures. For the experiments, the DBT molecules are embedded as an impurity into a thin AC film that acts as host matrix and protects the molecules from oxidation. As a result, DBT is much less affected by photobleaching, providing day-long intense fluorescence emission with up to ~ 1 MHz count rate [21]. The emitter's photostability is mainly limited by AC sublimation at room temperature. In our experiments, the DBT-doped AC crystals are fabricated through a simple spin-coating procedure. For this, an initial solution of DBT in toluene is prepared and later diluted in a solution of AC in diethylether with a concentration of 2.5 mg/ml. The DBT-doped AC crystals are then obtained by spin-coating a 20 μl droplet of this solution on a glass cover slip. During evaporation of the solvents formation of AC crystals takes place.

The AC crystals obtained from this procedure have clear-cut edges, average thickness of 20-100 nm (depending on concentrations and rotation velocity) and surface roughness on the order of 1 nm. Such characteristic is fundamental for integration with nanophotonic devices, as it allows efficient near-field coupling. Moreover, in order to minimize Gibbs free energy, DBT molecules are embedded in such a way that their molecular dipole moment is oriented parallel to the substrate, as was determined by back focal plane imaging [22].

Since the intersystem crossing to the DBT's triplet state is very unlikely ($1:10^4$), it has a quantum yield close to unity and hardly suffers from blinking. Moreover, when cooled down to liquid Helium temperatures, the transition to the ground state can exhibit a lifetime-limited linewidth [23–26]. Finally, the central emission wavelength around 785 nm promises a low damping rate, when plasmons in gold structures are excited.

DBT's optical stability and its protection inside the AC crystal open another attractive possibility, i.e., their nano-manipulation. It has been demonstrated that micromanipulators or scanning probes such as atomic force microscopes (AFM) can be utilized to control the position of single emitters actively with nanometer precision [7,27].

To demonstrate such a nano-manipulation with molecules, DBT-doped AC crystals were first prepared on a glass cover slip to allow investigation of fluorescence and topography at the same time. Figure 5 shows the setup used. A pulsed Ti:Sa laser ($\lambda \sim 750$ nm) with a repetition rate of 80 MHz excites DBT molecules. Fluorescence is detected by avalanche photo diodes (APDs) in a confocal configuration. Photon-correlation measurements are performed with a Hanbury Brown and Twiss (HBT) setup. An AFM directly on top of the microscope objective allows for topography measurements, fluorescence studies, and nano-manipulation of pre-selected AC crystals.

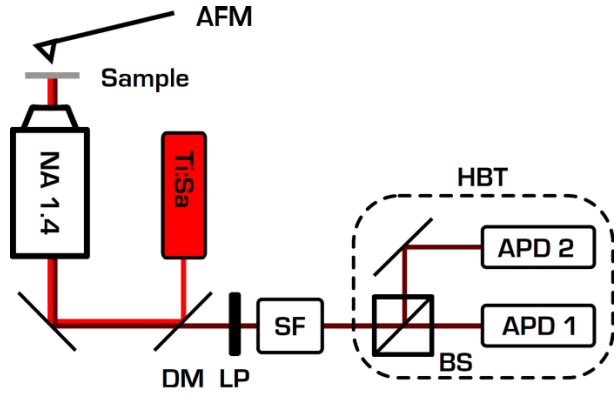

Fig. 5. Scheme of combined optical and atomic force microscope. A Ti:Sa Laser (λ ~ 750 nm), which can be run in either continuous wave (cw) or pulsed mode is used for excitation. An AFM on top of the optical microscope allows for correlated fluorescence and topography studies (in tapping mode) as well nano-manipulation and -structuring (in contact mode) of designated AC crystals. The detection (through a dichroic mirror (DM), a longpass filter (LP), a spatial filter (SF) and a beamsplitter (BS)) is equipped with a HBT setup that enables investigation of photon-correlations of fluorescence from single emitters.

Figure 6a) and b) show an example of nano-structuring and -manipulation of AC crystals. In between the two scans of the topography with the AFM (NT-MDT with NC-50 nanoworld, silicon tips) in tapping mode, the AFM was operated in contact mode to cut the AC crystal into two sub-micron-sized pieces. These fragments could be pushed several 100 nm across the substrate. In various experiments with different crystal thicknesses and DBT densities, we found that the stability of DBT fluorescence is not affected by this cutting and pushing procedure.

The results open the possibility to reduce the size of AC crystals in a controlled way, but additionally, to carve out small AC crystals with a certain number of DBT molecules inside. In order to demonstrate this, we performed the measurements shown in Figure 6c) to 6f). We started with a confocal mapping of a sample area and identified an AC crystal containing fluorescing DBT molecules as bright diffraction limited spot (Figure 6c). By measuring the normalized auto-correlation function $g^{(2)}(t)$ the number of DBT molecules in the AC crystal can be identified. For n identical emitters the equation $g^{(2)}(0) = 1-1/n$ holds [28]. For the AC crystal in Figure 6d) we can estimate n ~ 3-4 (see Fig. 6d)). Then, we applied the AFM cutting and nano-manipulation procedure on the same area and repeated the confocal scan. The result is shown in Fig. 6e). One can now clearly identify two spots identified as two separated AC crystals. Measurements of the autocorrelation functions on the two crystals (Fig. 6f)) reveal a number of probably 1 molecule ($g^{(2)}(0) < 0.5$ without background subtraction), and 2-3 molecules, respectively. The ability of precise carving and positioning of a single DBT molecule embedded in a single sub-micron AC is a very important step to fabricate a stable single quantum emitter as active element in a plasmonic non-linear device.

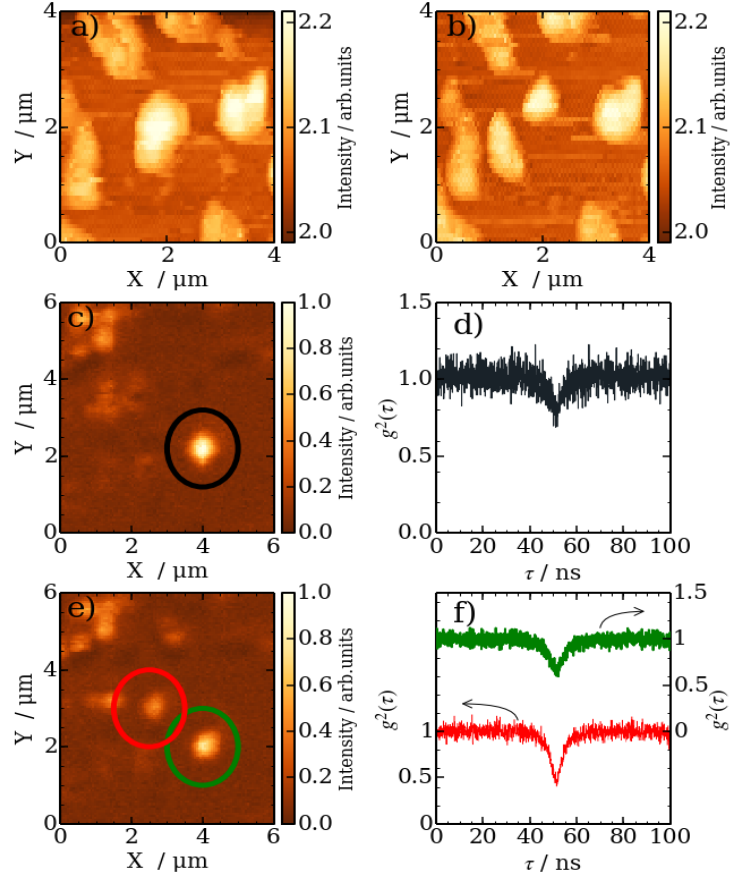

Fig. 6. a), b) AFM topography images of AC nanocrystals before and after nano-structuring with the AFM in contact mode. A large crystal was cut in two pieces, which were subsequently separated.
c), e) Confocal microscopy scans of a sample area with AC crystals containing DBT molecules before c) and after e) AFM cutting and nano-manipulation. d) and f) show measured autocorrelation functions of the encircled spots in. c) and e), respectively.

*iii) Assembling approaches*

The final requirement for fabricating a single-molecule non-linear element is to assemble the dielectric-plasmonic chip and the single molecules, as introduced in Sec. i) and Sec. ii), respectively. One approach could be to further develop the nano-manipulation technique described in Sec. ii) into a reliable pick-and-place method as demonstrated with color centers in nanodiamonds [29]. The non-transparent substrate of the plasmonic-photonic chip described in Sec. i), does not allow simultaneous optical investigation and nano-manipulation, as demonstrated on cover slips. To maintain a closed workflow the use of scanning near-field microscopy for simultaneous optical and topography inspection below the diffraction limit, or transparent substrates that allow optical measurements from below could be employed.

Here, we tested two other techniques: direct formation (during spincoating like described above) of AC crystals containing DBT molecules on the chip and transfer of pre-characterized AC crystals from glass substrates by direct contact. Figure 7 shows a result following the first approach. In an optical microscope image of the chip larger AC crystals of approximately 100 nm in thickness can be identified.

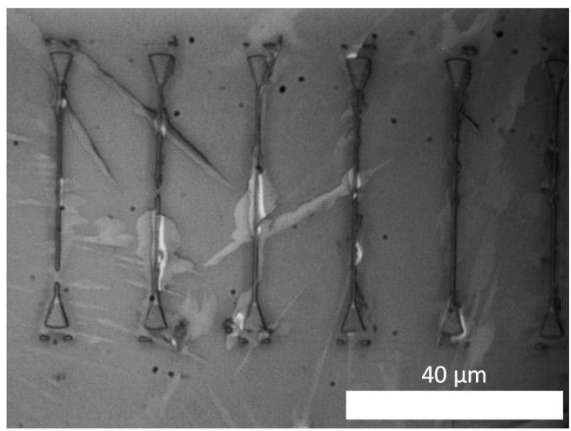

Fig. 7. Optical microscope image of AC crystals that formed during spin-coating directly on the chip on top of the waveguiding structures. Nucleation sites occur preferentially on rough structures, e.g. the Bragg-gratings or edges of the waveguides.

Although a broad range of AC crystal sizes were obtained, a truly self-assembled growth of very small AC crystals with only a few – ideally single – molecules could not be obtained yet. In a second approach a glass cover slip where AC crystals containing DBT molecules were formed again by spin-coating was brought in direct contact with the dielectric-plasmonic chip. In this first experiment of 'single molecule contact printing' no alignment between the chip and glass cover slip was employed. Therefore, the deposition of AC crystals on the chip was random. In order to analyze both the topography of the fabricated hybrid structure as well as the fluorescence from the DBT molecules we employed both confocal scattering as well as confocal fluorescence imaging. We used the setup described in Figure 5, but switched between the two situations where the Ti:Sa excitation laser was completely suppressed by filtering (confocal fluorescent mode) and where part of the excitation light reached the APD detector (confocal scattering mode). In scattering mode the dielectric and plasmonic waveguiding structures as well as the AC nanocrystals could be visualized (as depicted in Fig. 8a)). Switching back to confocal fluorescent mode the signal from the DBT molecules could be derived (Fig. 8b)).

It is apparent that in the first completed hybrid structure there was still a misalignment between the AC crystal and the plasmonic waveguide. This is expected in random fabrication process. However, Figure 8 clearly shows that our approach is capable to combine all necessary components for a single molecule non-linear element on one chip and to perform a structural and optical investigation on the single emitter level. Beyond the statistical approach [30] to look for a successful alignment of AC crystal and plasmonic coupler by mere chance, there are other options to improve the fabrication yield. One may fabricate markers on the chip as well as on the glass cover slip holding the AC crystals. Then an alignment of chip on cover slip during printing is possible. Another idea is to use subsequent AFM manipulation, once the AC crystals have been deposited, similar as in [31]. AFM manipulation may also be applied on the glass cover slip before printing in order to reduce the AC crystal size or to pushing smaller crystals.

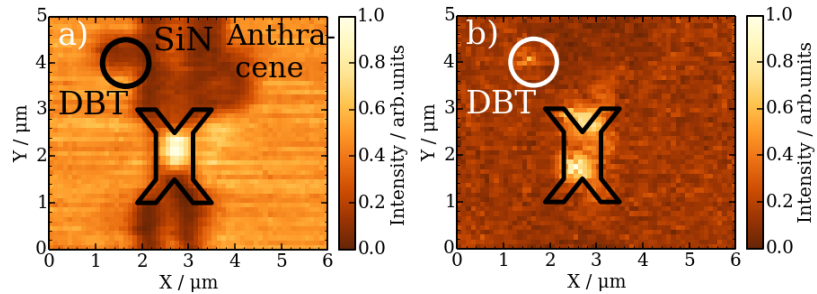

Fig. 8. a) Confocal scattering image of AC crystals containing DBT molecules close to a dielectric-plasmonic coupler after 'single molecule contact printing'. For clarity the shape of the gold plasmonic waveguide is highlighted. Increased reflection from gold can be well distinguished from increased scattering from dielectric waveguides and AC nanocrystals.
b) Confocal fluorescence image of the same area as in a). DBT fluorescence (encircled spot) is detected originating from the AC nanocrystal close to the dielectric waveguide. Additional luminescence from the gold (lower spot) is observed due to two-photon excitation by the Ti:Sa laser.

In conclusion we have demonstrated all necessary key components and technologies to realize an optical single molecule non-linear device based on the strongly enhanced interaction between single plasmons and individual DBT molecules. In particular, the on-chip photonic-plasmonic structures were fabricated and characterized in detail. Furthermore, we proved the usability of DBT in AC as emitter system and demonstrated a suitable nano-manipulation technique. Finally, a route for the alignment of the hybrid system consisting of the chip and AC crystals was outlined.

In future steps we will further develop the demonstrated methods to exploit the coupling of single emitters to the plasmonic waveguide structures. First, the extraction of single photons from the emitter will be investigated as a qualitative check to find perfect coupling performance and to build ultra-bright plasmon enhanced single photon sources. In a second step, single photons or attenuated laser-light will be coupled to the waveguides to investigate non-linear responses originating from the intense plasmon-molecule interaction. The presented components and technologies might find applications in other nano-optical experiments where complex on-chip structures or feasible, nanoscopic and spectrally narrow emitters are needed.

### Acknowledgments

Support by the German Research Foundation (DFG via SFB 951) is gratefully acknowledged.
We like to thank the team of JCMwave for supporting our numerical simulations.
CT and GM acknowledge support from the European Network of Excellence for Energy Efficiency, with the seed project OLEIT.